\documentclass{aa}
\usepackage{txfonts}
\usepackage{natbib}
\usepackage{graphicx}
\usepackage{url}

\bibpunct{(}{)}{;}{a}{}{,}
\def\be{\begin{equation}}
\def\ee{\end{equation}}
\def\bea{\begin{eqnarray}}
\def\eea{\end{eqnarray}}
\titlerunning{The initial Lorentz factors from early Swift-XRT data}

\begin{document}
\title{The initial Lorentz factors of fireballs inferred from the early X-ray data of SWIFT GRBs}
\author{Rong-Rong Xue\inst{1,2}, Yi-Zhong Fan\inst{3,1}, Da-Ming Wei\inst{1,4}}

\institute{Purple Mountain Observatory, Chinese Academy of
Sciences, Nanjing 210008, China \and  Graduate School, Chinese
Academy of Sciences, Beijing, 100012, China \and Niels Bohr
International Academy, Niels Bohr Institute, University of
Copenhagen, Blegdamsvej 17, DK-2100 Copenhagen, Denmark \and Joint
Center for Particle Nuclear Physics and Cosmology of Purple
Mountain Observatory - Nanjing University, Nanjing 210008, China
}\mail{Y.Z. Fan (yizhong@nbi.dk)}

\date{Received / Accepted}

\abstract{}{We intend to determine the type of circumburst medium
and measure directly the initial Lorentz factor $\Gamma_0$ of GRB
outflows.} {If the early X-ray afterglow lightcurve has a peak and
the whole profile across the peak is consistent with the standard
external shock model, the early rise profile of light curves can
be used to differentiate whether the burst was born in
interstellar medium (ISM) or in stellar wind.  In the thin shell
case, related to a sub-relativistic reverse shock, the peak time
occurring after the end of the prompt emission, can be used to
derive an accurate $\Gamma_0$ , especially for the ISM case. The
afterglow lightcurves for a flat electron spectrum $1<p<2$ have
been derived analytically.} {In our GRB sample, we obtain
$\Gamma_0 \sim 300$ for the bursts born in ISM. We did not find
any good case for bursts born in stellar wind and behaving as a
thin shell that can be used to constrain $\Gamma_0$ reliably.}{}


\maketitle

\section{Introduction}
Gamma-ray bursters  are among the most mysterious celestial
objects and have attracted people since its first detection in
1967 \citep{Kle73}. The time variability of pulses, as short as
millisecond, limits this event to a object of the stellar scale.
The random occurrence and also the short time-duration of this
kind of event lead to difficulties in detection. The dark era of
research on Gamma-ray bursts (GRBs) lasts until the release of
X-ray afterglow data of GRB 970228, confirming GRBs at the
cosmological distances \citep{Costa97}.

The power-law decay of multi-waveband afterglows of many GRBs are
consistent with the standard external shock model
~\citep{Waxman97,Wijers97}. However, the multi-waveband afterglows
are usually monitored several hours after the burst trigger. The
late afterglow, independent on the initial values of the fireball,
can not provide us information about the fireball characteristics.
The \emph{Swift} satellite \citep{Gehrels04}, thanks to its rapid
response time and accurate localization, X-ray Telescope (XRT),
Ultra-Violet Telescope (UVOT) on board and other ground-based
telescopes can slew to GRB within tens seconds and then begin
observations. The early afterglow data released in Swift era provide
us an opportunity  to study properties of fireballs, e.g., the
initial Lorentz factor of the fireball.

The fireball is expected to be a highly relativistic ejection from
the central engine to avoid the "compact problem"
\citep{SP90,Lithwick01}. After the radiation-dominated
acceleration phase, the fireball goes into a matter-dominated
phase when the fireball is no long accelerated. The fireball keeps
an approximately invariable velocity until it sweeps up
considerable mass of ambient medium. We call this episode as the
coasting phase ~\citep{Piran93}. Though the profile of the early
afterglow exhibits quite different from burst to burst and also
from X-ray to infrared (IR) band, the peaks in the early afterglow
light curves may indicate the arrival of the deceleration radius
$(R_{\rm d})$ in some GRBs. For example, \citet{Molinari07}
attributed the peaks in near-Infrared afterglows of GRB 060418 and
GRB 060607A to the end of the coasting phase and determined the
initial Lorentz factors of the fireballs (see also Jin \& Fan
2007). Recently, Oates et al. (2009) analyzed the early afterglows
of {\it Swift}-UVOT data and measured the initial Lorentz factor
of GRBs for those showing an early power-law increase in flux.

Different from these works, now we use the early X-ray data of {\it
Swift} GRBs to constrain the initial Lorentz factors ($\Gamma_0$).
As a probe of $\Gamma_0$, the X-ray data is better than the optical
data for the following reasons: (1) In the standard fireball model,
the X-ray emission decays with time quickly after the outflow has
got decelerated, independent of the profile of the medium
surrounding the progenitor \citep{Fan05}. This is because usually
both the typical frequencies of the forward shock (FS) and the
reverse shock (RS) emission are below the X-ray band (see Tab.1 for
the light curves). The optical emission, however, will increase
until the typical synchrotron frequency of the FS drops below the
optical band \citep{SPN98}. (2) In the thin shell case that is of
our interest, usually the RS X-ray emission is not strong enough to
outshine the FS emission component. The origin of the X-ray peak can
thus be reliably established. (3) X-ray afterglows are hardly
influenced by the self-absorption effect and dust extinction,
different from the emission at lower frequencies. One disadvantage
of our method is that the early X-ray emission of most GRBs have
been polluted by the delayed flares \citep{Fal08} powered by the
prolonged activity of the central engine. Fortunately, the X-ray
flares usually have a decline as steep as $t^{-(3\sim 10)}$, which
is significantly sharper than what the fireball model predicts. So
one can distinguish the peak of FS emission from the peak of flare
in X-ray band convincingly.

\section{The early X-ray afterglow emission}
A very bright optical flash has been detected in GRB 990123
\citep{Akerlof99}. The most widely discussed interpretation is the
external RS model (Sari \& Piran 1999; M\'esz\'aros \& Rees 1999
however, it can also be produced by the internal shock model, e.g.
Wei 2007). Since then, the RS emission in optical band has been
extensively investigated (see Zhang 2007 for a review). However, the
RS X-ray emission has just been calculated by a few authors
\citep{Fan05,Zou05}. In this work we focus on the profile of early
X-ray afterglow light curves in different cases.

In the Fermi acceleration process, the power-law index of shocked
electrons $p>2$ is resulted \citep{Gallant02}. It has been taken
as the standard scenario and has been widely used in the afterglow
calculation \citep{SPN98,CL00}. Some afterglow modelings
\citep{Baha01,DC01}, however, favor a flat electron spectrum
$1<p<2$, for which a reliable estimate of the afterglow emission
is still unavailable. We'll discuss such a scenario in section 2.2
in some detail.

\subsection{The case of $p>2$}
Firstly, we discuss the ``thin shell case'' (for which
$t_\times>T_{90}$, where $t_\times$ is the crossing time of RS and
$T_{90}$ is the duration of the burst), referring to a
sub-relativistic RS. Assuming typical parameters (e.g., the
fraction of the shock energy given to the electrons
$\epsilon_e=0.1$, the fraction given to the magnetic field
$\epsilon_B=0.01$, the total energy of the fireball $E=10^{52}{\rm
erg}$ and $p=2.3$), we have the typical synchrotron radiation
frequency and the cooling frequency of the FS emission
$\nu_m^f=2\times 10^{16}{\rm Hz}~(\rm t/100s)^{-3/2}$,
$\nu_c^f=8\times 10^{16}{\rm Hz}~ n_0^{-1}(\rm t/100s)^{-1/2}$ for
bursts born in ISM \citep{SPN98} and $\nu_m^f=5\times10^{16}{\rm
Hz}~(\rm t/100s)^{-3/2}$, $\nu_c^f=4.8\times10^{12}{\rm Hz}~(\rm
t/100s)^{1/2}A_*^{-2}$ born in wind medium \citep{CL00}, where $n$
and $A_*$ are two medium parameters--~$n$ indicates the density of
ISM, in unit of $cm^{-3}$, while $A_*$ is the dimensionless
parameter of stellar wind environment. $t$ indicates the observed
time postburst. The convention $Q_x=Q/10^x$ has been adopted in
this paper, in units of cgs.

Combined with $\nu{^f}_{ m}(t_\times)>\nu{^r}_{m}(t_\times)$,
$\nu{^f}_{ c}(t_\times)\approx\nu{^r}_{c}(t_\times)$, we find that
both $\nu_m$ and $\nu_c$ of FS and RS emission (marked by the
subscripts $f$ and $r$ respectively) are below the X-ray band
$\nu_{\rm x}\sim 10^{17}$~Hz, assuming typical parameters.
 If the shock parameters are similar for the FS and
the RS, the flux contrast between the RS and FS X-ray emission is
thus\footnote{ This expression is valid in the ``thin shell" case.
In the ``thick shell" case, such an expression is an upper limit
since the total number of the RS electrons is less than $\Gamma_0$
times that of the FS electrons (see
eq.(\ref{eq:flux_contrast_thick})). } \citep{Fan05} \bea {F_{\rm
\nu_{\rm _X}}^{\rm r}(t_\times) \over F_{\rm \rm \nu_{\rm
_X}}^{\rm f}(t_\times)}
& \approx & \Gamma_0 ({\gamma_{34,\times}-1 \over
\Gamma_\times-1})^{\rm p-1},\label{eq:Fnu}\eea where the subscript
$\times$ represents the parameters measured at $t_\times$.
$\Gamma_0$ and $\Gamma_\times$ represent the initial Lorentz
factor and the instant Lorentz factor at $t_\times$, respectively.
The corresponding relative Lorentz factor is
$\gamma_{34,\times}\approx
(\Gamma_0/\Gamma_\times+\Gamma_\times/\Gamma_0)/2$.

In the so-called ``thin shell case" , $\Gamma_\times \approx
\Gamma_0/2$ and $\gamma_{34,\times} \approx 1.25$. For a typical
$p\sim2.3$, Eq.(\ref{eq:Fnu}) gives
\begin{equation}
{F_{\rm \nu_{\rm _X}}^{\rm r}(t_\times) \over F_{\rm \rm \nu_{\rm
_X}}^{\rm f}(t_\times)} \approx 2^{1-p} \Gamma_0^{2-p} \sim {\cal
O}(0.1),\label{eq:flux_1}
\end{equation}
for $\Gamma_0 \sim~{\rm a~few~\times 100}$, {\it which suggests
that the RS X-ray emission can be ignored.} {This conclusion is
unchanged if $\nu_{\rm m}^f$ is actually above $\nu_{_{\rm X}}$
because in such a case the FS X-ray emission would be stronger.}

In the ``thick shell case" (for which $t_\times \sim T_{90}$),
particularly for a relativistic RS satisfying $\gamma_{34,\times}-1
\approx \Gamma_0/2\Gamma_\times$, we have
\begin{equation}
{F_{\rm \nu_{\rm _X}}^{\rm r}(t_\times) \over F_{\rm \rm \nu_{\rm
_X}}^{\rm f}(t_\times)} < 2 (\gamma_{34,\times}-1)^{p}
\Gamma_\times^{2-p}.\label{eq:flux_contrast_thick}
\end{equation}
As a result, the RS X-ray emission may be able to outshine the FS
component.

Thin shell case: The RS X-ray emission is unimportant, so the
afterglow is dominated by the FS component. The bulk Lorentz
factor $\Gamma$ is  nearly a constant in this coasting phase
($t<t_\times$). We then have $\nu_m^{f} \propto t^{-k/2}$,
$\nu_c^{f} \propto t^{3k/2-2}$, and the maximal specific flux
$F_{\rm \nu,max}^{f} \propto t^{3(1-k/2)}$ ($k=0$ for ISM and
$k=2$ for the stellar wind). So the FS X-ray emission evolves as
$F_{\nu_{\rm _X}}^{\rm f}\propto t^{2-(2+p)k/4}$ with time. In the
case of ISM, $F_{\nu_{\rm _X}}^{\rm f}\propto t^{2}$, increasing
with time quickly, while in the wind case, $F_{\nu_{\rm _X}}^{\rm
f}\propto t^{(2-p)/2}$, decreasing with time slowly.

Thick shell case: In the case of ISM, we have $\Gamma \propto
t^{-1/4}$, $\nu_m^{\rm r} \propto t^0$, $\nu_c^{\rm r} \propto
t^{-1}$, $F_{\nu,\rm max}^{r} \propto t^{1/2}$, and then derive
$F_{\nu_{\rm _X}}^{\rm r}\propto t^{0}$ (see also Kobayashi 2000)
when $t<t_\times$. The simultaneous FS X-ray emission is
$F_{\nu_{\rm _X}}^{\rm f}\propto t^{(2-p)/2}$. In the wind case,
both the FS and the RS emission decrease with time slowly as
$t^{(2-p)/2}$ when $t<t_\times$ (see also Fan \& Wei 2005).

For~ $t>t_\times$, it is well known that $F_{\nu_{\rm _X}}^{\rm
f}\propto t^{(2-3p)/4}$, independent of the type of circumburst
medium.

 Please note that in our above analysis, we assume that both
$\nu_m$ and $\nu_c$ of FS and RS are well below the XRT band. More
general  results have been summarized in Tab.1. One can see that
$F_{\nu_{\rm _X}}^{\rm f}\propto t^{-1/4}$ is also possible but
only for a fast cooling forward shock. Its spectrum should be
$F_\nu \propto \nu^{-1/2}$, which can  be distinguished from the
shallow decline predicted for the FS and RS emission in the wind
case at a time $t<t_{\rm p}$.

\begin{table}
\caption{The temporal behavior of the X-ray afterglow lightcurves in
the case of $p>2$.}
\begin{tabular}{llll}
\hline

& emission regime & ISM  &  wind  \\
\hline
thin shell case (FS)& & & \\
\hline $t<t_{\rm p}$ &$\nu_{_{\rm X}}>\max\{\nu_c^f,~\nu_m^f\}$  &
$t^2$  &   $t^{(2-p)/2}$\\
$t<t_{\rm p}$ & $\nu_c^f<\nu_{_{\rm X}}<\nu_m^f$  &  $t^2$  &  $t^{1/2}$  \\
$t<t_{\rm p}$ & $\nu_m^f<\nu_{_{\rm X}}<\nu_c^f$  &  $t^3$  &   $t^{-(p-1)/2}$ \\
\hline {thin shell case (RS)}& & &\\
\hline $t<t_{\rm p}$ &$\nu_m^r<\nu_c^r<\nu_{_{\rm X}}$ &$t^{2p-1}$  &$t^{-(p-2)/2}$\\
  $t<t_{\rm p}$ &$\nu_m^r<\nu_{_{\rm X}}<\nu_c^r$ &$t^{2p}$ &$t^{-(p-1)/2}$ \\
 \hline
 thick shell case (RS) & & & \\
\hline
 $t<t_{\rm p}$ & $\nu_{_{\rm X}}>\max\{\nu_c^r,~\nu_m^r\}$ &  $t^0$  &   $t^{(2-p)/2}$ \\
$t<t_{\rm p}$ & $\nu_c^r<\nu_{_{\rm X}}<\nu_m^r$  &  $t^0$  &  $t^{1/2}$  \\
$t<t_{\rm p}$ & $\nu_m^r<\nu_{_{\rm X}}<\nu_c^r$  &  $t^{1/2}$  &   $t^{-(p-1)/2}$ \\
\hline
Late FS emission & & & \\
\hline $t>t_{\rm p}$ &$\nu_{_{\rm X}}>\max\{\nu_c^f,~\nu_m^f\}$  &
$t^{(2-3p)/4}$  &   $t^{(2-3p)/4}$\\
$t>t_{\rm p}$ & $\nu_c^f<\nu_{_{\rm X}}<\nu_m^f$  &  $t^{-1/4}$  &  $t^{-1/4}$  \\
$t>t_{\rm p}$ & $\nu_m^f<\nu_{_{\rm X}}<\nu_c^f$  &  $t^{3(1-p)/4}$  &   $t^{(1-3p)/4}$ \\
\hline
Late FS emission (jet) & & & \\
\hline $t>t_{\rm j}$ &$\nu_{_{\rm X}}>\max\{\nu_c^f,~\nu_m^f\}$  &
$t^{-p}$  &   $t^{-p}$\\
$t>t_{\rm j}$ & $\nu_c^f<\nu_{_{\rm X}}<\nu_m^f$  &  $t^{-1}$  &  $t^{-1}$  \\
$t>t_{\rm j}$ & $\nu_m^f<\nu_{_{\rm X}}<\nu_c^f$  &  $t^{-p}$  &   $t^{-p}$ \\
\hline
\end{tabular}
{Note --- Observationally the crossing time $t_\times$ marks the
beginning of the late sharp decline, in this work we denote such a
timescale by $t_{\rm p}$. $t_{\rm j}$ is the jet break time.}
\label{Tab:sum}

\end{table}



\subsection{The case of $1<p<2$}
With the shock jump conditions, \cite{Baha01} derived a minimum
Lorentz factor ($\gamma_{\rm m}$) of the electrons that depends on
the maximal one ($\gamma_{\rm M}$), i.e., $\gamma_{\rm m} \approx
\{(2-p)m_{\rm p}\epsilon_{\rm e}\Gamma_{\rm sh}\gamma_{\rm
M}^{p-2}/[(p-1)m_{\rm e}]\}^{1/(p-1)}$, where  $\Gamma_{\rm sh}$
is the Lorentz factor of the shock, $m_{\rm p}$ and $m_{\rm e}$
are the rest mass of protons and electrons, respectively. However,
in reality, particle acceleration proceeds from low to high
energy. $\gamma_{\rm m}$ and $\gamma_{\rm M}$ should be determined
by the first shock-crossing and  by radiative losses or escape
from the acceleration region, respectively. Hence $\gamma_{\rm m}$
should have no ``knowledge" of $\gamma_{\rm M}$ (Panaitescu 2006,
private communication). Motivated by the above arguments, in this
work we assume that $\gamma_{\rm m}\approx (\epsilon_{\rm e}/{\rm
f}(p))(\Gamma_{\rm sh}-1)m_{\rm p}/m_{\rm e}$,
where f($p$) being a function of $p$. Such  treatment requires
that only a small fraction ${\cal R}$ of the upstream material has
been accelerated otherwise the energy momentum conservation law
will be violated.

Assuming that the shock-accelerated electrons have a power-law
energy distribution ${dn}/d{\gamma_{\rm e}} \propto
 ({\gamma_{\rm e}}-1)^{-p}$ for $\gamma_{\rm m} \leq {\gamma_{\rm e}} \leq {\gamma}_{\rm M}$,
 with the shock jump conditions that $\int^{\gamma_{\rm
M}}_{\gamma_{\rm m}} ({dn}/d{\gamma_{\rm e}}) d{\gamma_{\rm
e}}=4{\cal R} \Gamma_{\rm sh}n_{\rm u}$ and $\int^{\gamma_{\rm
M}}_{\gamma_{\rm m}} ({\gamma_{\rm e}}-1) m_{\rm e} c^2
({dn}/d{\gamma_{\rm e}}) d{\gamma_{\rm e}}=4\Gamma_{\rm
sh}(\Gamma_{\rm sh}-1)\epsilon_{\rm e} n_{\rm u}m_{\rm p}c^2$, we
have
\begin{eqnarray}
{\cal R} & \approx & {2-p \over p-1} \epsilon_{\rm
e}({\epsilon_{\rm e}\over f(p)})^{1-p}({m_{\rm p} \over m_{\rm
e}})^{2-p}(\Gamma_{\rm
sh}-1)^{2-p}\gamma_{\rm M}^{p-2}\nonumber\\
&\propto & (\Gamma_{\rm sh}-1)^{2-p} B^{1-p/2}. \label{eq:f}
\end{eqnarray}
In this work we also assume that the maximal Lorentz factor is
limited by the synchrotron losses and is given by \citep{wc96}
\begin{equation}
\gamma_{\rm M}\approx 4\times 10^7 {B}^{-1/2},
\end{equation}
$B=\sqrt{32\pi \varepsilon_{\rm B}\Gamma_{\rm sh}(\Gamma_{\rm
sh}-1)n_{\rm u} m_{\rm p}c^2}$ is the magnetic field strength of
the shock region,  $n_{\rm u}$ is the number density of the
upstream medium measured in its rest frame and $c$ is the speed
velocity of light.

The afterglow lightcurves in the case of $1<p<2$ are different from
those presented in Tab.\ref{Tab:sum} by a factor of ${\cal R}$ given
below.

Assuming the X-band is above $\max\{\nu_{\rm c}^{\rm r},~\nu_{\rm
m}^{\rm r},~\nu_{\rm c}^{\rm f},~\nu_{\rm m}^{\rm f}\}$, the flux
contrast between the RS and the FS emission can be estimated by
(for a thin shell)
\begin{equation}
{F_{\nu_{\rm x}}^{\rm r}(t_\times) \over F_{\nu_{\rm x}}^{\rm
f}(t_\times)}\approx \Gamma_0 ({\gamma_{34,\times}-1 \over
\Gamma_\times-1})^{p-1}({\gamma_{34,\times}-1 \over
\Gamma_\times-1})^{2-p} \sim {1\over 2}.
\end{equation}
As a result, in the thin shell case, the RS X-ray emission is
usually outshone by the FS X-ray radiation for both a flat electron
spectrum ($1<p<2$) and a standard electron spectrum ($p>2$).

For the FS before getting decelerated, we have $\Gamma_{\rm sh} \sim
t^{0}$, $B \propto R^{-k/2}\propto t^{-k/2}$, and ${\cal R}\propto
t^{-k(2-p)/4}$. For the GRBs born in ISM: In the case of a thin
shell, for the RS we have $\Gamma_{\rm sh}-1 \propto t^2$
\citep{Fan02}, $B \propto t^{0}$, and ${\cal R} \propto t^{2(2-p)}$;
In the case of a thick shell, for the RS we have $\Gamma_{\rm sh}-1
\propto t^{-1/4}$ \citep{Kob00}, $B \propto t^{-(2+3k)/8}$, and
${\cal R} \propto t^{-(2-p)(6+k)/16}$. For the GRBs born in wind:
for the RS we have $\Gamma_{\rm sh}-1 \propto t^{0}$, $B \propto
t^{-1}$, and ${\cal R} \propto t^{-(2-p)/2}$.

For the FS taking a Blandford-McKee profile, we have $\Gamma_{\rm
sh} \sim \Gamma \propto R^{(k-3)/2} \propto t^{(k-3)/(8-2k)}$, $B
\propto \Gamma_{\rm sh}R^{-k/2}\propto R^{-3/2}\propto
t^{-3/(8-2k)}$, and ${\cal R}\propto t^{(2k-9)(2-p)\over 4(4-k)}$.
As long as the edge of the ejecta is visible, the sideways expansion
may be so important that can not be ignored, with which the dynamics
is governed by $\Gamma \propto t^{-1/2}$ \citep{Rhoads99}. We then
have ${\cal R} \propto t^{-3(2-p)/4}$, independent of $k$.

The detailed lightcurves have been summarized in Tab.\ref{Tab:sum2}.

\begin{table}
\caption{The temporal behavior of the X-ray afterglow lightcurves in
the case of $1<p<2$.}
\begin{tabular}{llll}
\hline

& emission regime & ISM  &  wind  \\
\hline
thin shell case (FS)& & & \\
\hline $t<t_{\rm p}$ &$\nu_{_{\rm X}}>\max\{\nu_c^f,~\nu_m^f\}$  &
$t^2$  &   $t^{0}$\\
$t<t_{\rm p}$ & $\nu_c^f<\nu_{_{\rm X}}<\nu_m^f$  &  $t^2$  &  $t^{(p-1)/2}$  \\
$t<t_{\rm p}$ & $\nu_m^f<\nu_{_{\rm X}}<\nu_c^f$  &  $t^3$  &   $t^{-1/2}$ \\
\hline {thin shell case (RS)}& & &\\
\hline $t<t_{\rm p}$ &$\nu_m^r<\nu_c^r<\nu_{_{\rm X}}$ &$t^{3}$  &$t^{0}$\\
  $t<t_{\rm p}$ &$\nu_m^r<\nu_{_{\rm X}}<\nu_c^r$ &$t^{4}$ &$t^{-1/2}$ \\
 \hline
 thick shell case (RS) & & & \\
\hline
 $t<t_{\rm p}$ & $\nu_{_{\rm X}}>\max\{\nu_c^r,~\nu_m^r\}$ &  $t^{-3(2-p)/8}$  &   $t^{0}$ \\
$t<t_{\rm p}$ & $\nu_c^r<\nu_{_{\rm X}}<\nu_m^r$  &  $t^{-3(2-p)/8}$  &  $t^{(p-1)/2}$  \\
$t<t_{\rm p}$ & $\nu_m^r<\nu_{_{\rm X}}<\nu_c^r$  &  $t^{(3p-2)/8}$  &   $t^{-1/2}$ \\
\hline
Late FS emission & & & \\
\hline $t>t_{\rm p}$ &$\nu_{_{\rm X}}>\max\{\nu_c^f,~\nu_m^f\}$  &
$t^{-(10+3p)/16}$  &   ${t^{-(6+p)/8}}$ \\
$t>t_{\rm p}$ & $\nu_c^f<\nu_{_{\rm X}}<\nu_m^f$  &  $t^{-(22-9p)/16}$  &  ${t^{-(12-5p)/8}}$  \\
$t>t_{\rm p}$ & $\nu_m^f<\nu_{_{\rm X}}<\nu_c^f$  &  ${t^{-(6+3p)/16}}$  &   ${t^{-(8+p)/8}}$ \\
\hline
Late FS emission (jet) & & & \\
\hline $t>t_{\rm j}$ &$\nu_{_{\rm X}}>\max\{\nu_c^f,~\nu_m^f\}$  &
$t^{-(p+6)/4}$  &   $t^{-(p+6)/4}$\\
$t>t_{\rm j}$ & $\nu_c^f<\nu_{_{\rm X}}<\nu_m^f$  &  ${t^{(3p-10)/4}}$  &  ${t^{(3p-10)/4}}$  \\
$t>t_{\rm j}$ & $\nu_m^f<\nu_{_{\rm X}}<\nu_c^f$  &  $t^{-(p+6)/4}$  &   $t^{-(p+6)/4}$ \\
\hline
\end{tabular}
{Note --- Observationally the crossing time $t_\times$ marks the
beginning of the late sharp decline, in this work we denote such a
timescale by $t_{\rm p}$. $t_{\rm j}$ is the jet break time.}
\label{Tab:sum2}
\end{table}

\section{case studies}
In the thick shell case, the RS disappears almost simultaneously
with the prompt X-ray and $\gamma-$ray emission. As a result, the RS
X-ray emission may be outshone by the prompt emission component.
Moreover, for a thick shell, we need a self-consistent modeling of
the FS and RS emission to get $\gamma_{34,\times}$ and then
$\Gamma_0$. In such a process, quite a few free parameters are
introduced and the constraint on $\Gamma_0$ is rather uncertain.
{\it In this work, we only focus on the thin shell case with $t_{\rm
p}=t_{\times}>T_{90}$.}

As summarized in Tab.\ref{Tab:sum} and Tab.\ref{Tab:sum2}, in the
thin shell case, the outflow expanding into the ISM will give rise
to an increase not shallower than $t^2$, while expanding into the
wind can not account for an increase steeper than $t^{1/2}$.
Therefore we can judge whether the GRB was born in ISM or wind
medium according to the sharpness of flux increase. One can also
speculate that the X-ray data for the GRBs born in wind is not a
very good probe of the initial Lorentz factor of the ejecta due to
the lack of a distinguished peak for typical parameters that give
rise to $\nu_{\rm X}>\max\{\nu_{m},\nu_c\}$.


The number of GRBs recorded by XRT exceeds 160 till August 1st,
2008. Most of them, however, play no role in constraining
$\Gamma_0$. For our purpose, the light curves have to be
characterized by: (i) There is a distinguished/single peak. (ii)
Across the peak, a smooth transition to a single power-law decay is
followed, and the whole profile must be consistent with the standard
afterglow model \citep{SPN98,CL00}.

We checked all the {\it Swift} X-ray afterglows and find 7 GRBs
whose early peaks may be attributed to the arrival of $R_{\rm d}$
and the post-peak temporal behaviors of these bursts are
consistent with the standard afterglow model. All these bursts
except GRB 051111 have the X-ray afterglow peaks much later than
the end of the prompt emission and thus are in thin shell case. A
good example is that of GRB 080319C (see Fig.1). We divide our GRB
sample, in total 6 events, into two groups: 4 were born in ISM and
2 that might be born in wind, based on the rise behavior of the
early X-ray afterglow lightcurve.

\begin{figure}
\begin{center}
\includegraphics[width=250pt]{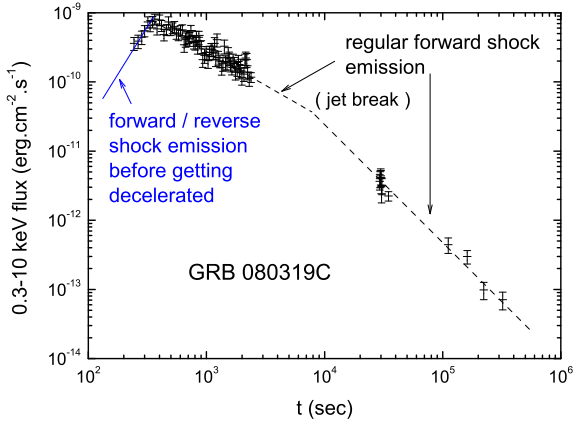}
\end{center}
\caption{The X-ray afterglow light curve of GRB 080319C and its
interpretation. The XRT data are from
{http://www.swift.ac.uk/xrt\_curves/00306778/} \citep{Evan07}.}
\end{figure}

\subsection{ISM case}

{\bf GRB 060801:} The BAT light curve showed a short burst with a
duration of about 0.5 second. The candidate host galaxy has a
redshift of 1.131 as given by ~\citet{Cucchiara06}. The XRT began
taking data 63 seconds after the BAT trigger  ~\citep{Racusin06a}.
The flux increased with time in form of $\sim t^{2.3}$ and peaked
$\sim$ 110s after the BAT trigger, followed by a decay of $ ~\sim
t^{ -1.19\pm0.16}$ with time to $\sim$ 400s from the
trigger~\citep{Racusin06b}. The XRT spectrum is $F_\nu \propto
\nu^{-0.67\pm 0.2}$. The decline and the spectrum are consistent
with the fireball afterglow model supposing that $p\sim 2.3$ and
$\nu_{\rm m}^{f}<\nu_{_{\rm X}}<\nu_{\rm c}^{f}$ (see
Tab.\ref{Tab:sum}). It is straightforward to show that for
$\nu_{\rm m}^{f}<\nu_{_{\rm X}}<\nu_{\rm c}^{f}$, the FS X-ray
emission still dominates over the RS X-ray component.

{\bf GRB 060926:} The peak occurred around 430s after the BAT
trigger. Before and after the peak, the flux evolved as $\sim
t^{2.0}$ and $\sim t^{-1.4}$, respectively. Though
~\citet{Holland06} considered this peak as a flare, the post-peak
flux decayed temporally same as the late time afterglows.
~\citet{Elia06} claimed a redshift of z=3.20 based on a damped
Ly$\alpha$ line and various absorption features. The XRT spectrum
is $F_\nu \propto \nu^{-1.1\pm 0.2}$. The decline and the spectrum
are consistent with the fireball afterglow model supposing that
$p\sim 2.4$ and $\nu_{_{\rm X}}>\max\{\nu_{\rm c}^{f}, \nu_{\rm
m}^{\rm f}\}$ (see Tab.\ref{Tab:sum}).

{\bf GRB 080319C:} The XRT began observing the burst 224 seconds
after the BAT trigger ~\citep{Pagani08}. The redshift z=1.95 was
found from absorption lines of the afterglow spectrum
~\citep{Wiersema08}. As shown in Fig.1, a quick increase in flux
$\sim t^{2.4}$ is obvious up to $\sim 360$s after the GRB trigger.
After the peak time, the late afterglow exhibited a broken
power-law -- a relatively shallow decline $\sim t^{-0.86}$
followed by a $\sim t^{-1.87}$ profile. We attribute the late
steep decline to the jet effect of this GRB. With a flat electron
spectrum $p\sim 1.5$, both the temporal behavior of lightcurves in
the two late episodes and the X-ray spectrum $F_{\nu}\propto
\nu^{-0.74\pm 0.06}$ can be well reproduced providing that
$\nu_{_{\rm X}}>\max\{\nu_{\rm c}^{f}, \nu_{\rm m}^{\rm f}\}$ (see
Tab.\ref{Tab:sum2}).

{\bf GRB 080413B}: The BAT light curve showed a single peaked
structure with a duration of about 3 seconds \citep{Stamaticko08}.
The early XRT data recorded a steep increase in flux, peaking
around 167 s from the trigger. The X-ray light curve is well
fitted by a simple power-law, with a decay slope of 0.88 $\pm$
0.06 from 167 s to $10^4$ s after the burst trigger
\citep{Troja08}. Based on the detection of numerous absorption
features, including Fe II, Mg II and Mg I lines,
\citet{Vreeswijk08} inferred a redshift z = 1.10. The XRT spectrum
is $F_\nu \propto \nu^{-1.05\pm 0.1}$. The decline and the
spectrum are roughly consistent with the fireball afterglow model
supposing that $p\sim 2$ and $\nu_{_{\rm X}}>\max\{\nu_{\rm
c}^{f}, \nu_{\rm m}^{\rm f}\}$ (see Tab.\ref{Tab:sum}).

\subsection{wind case?}

{\bf GRB 080307:} In the first orbit of the XRT data, the emission
rose slowly, $\sim t^{0.6}$, peaking at $\sim 240$s after the BAT
trigger, after which the decay can be modeled with a single
power-law with the temporal index $\alpha=1.83\pm 0.08$ up to
several ks after the burst trigger, though there was an abnormal
flat in flux later ~\citep{Page08}. The X-ray spectrum is $F_\nu
\propto \nu^{-0.74\pm 0.22}$. As shown in Tab.\ref{Tab:sum}, in the
thin shell case, for $\nu_{\rm c}^{f}<\nu_{_{\rm X}}< \nu_{\rm
m}^{\rm f}$, the FS emission of an ejecta expanding into the stellar
wind can give rise to an initial rise $t^{0.5}$, roughly consistent
with the data. However, this model gives a decline $t^{-1/4}$ after
the X-ray peak, deviating from the data significantly. {\it So this
event won't be included in our $\Gamma_0$ constraint.}

{\bf GRB 080409:} The XRT began observing this burst 84 seconds
after the BAT trigger. The light curve showed an initial increase
with a power-law slope of $\sim0.6$ and entered a peak at $\sim509$s
after the burst onset. The light curve then turned over to decay
with a power-law slope of 0.89$\pm$0.09 ~\citep{Holland08}. The XRT
spectrum is $F_\nu \propto \nu^{-1.2\pm 0.6}$. Similar to GRB
080307, an initial rise $t^{0.6}$ may be the FS emission of an
ejecta born in wind. However, this model gives a decline $t^{-1/4}$
after the X-ray peak unless we have seen the edge of the ejecta for
which the decline can be as steep as $t^{-1}$ (see
Tab.\ref{Tab:sum}). {\it Again, we won't include it in the following
$\Gamma_0$ constraint.}\\

\subsection{determining the initial Lorentz factor}
In our GRB sample, the peaks appearing in the early X-ray afterglow
light curves are related to the position of the outflow $R_{\rm d}$
from the central engine. At this time, the swept medium has an
energy comparable to that of the GRB outflow and the instant Lorentz
factor at $R_{\rm d}$ is about half of the initial one. Based on the
assumption above, we can measure the initial Lorentz factor of GRBs
in ISM case \citep{BM76}:
\begin{equation}
\Gamma_0  = [\frac{24E_{\gamma}(1+z)^3}{\pi n m_p c^5\eta
t_p^3}]^{1/8},
\end{equation}
where $E_{\gamma}$ is the isotropic energy of prompt gamma-ray
emission. Here, we take the radiation efficiency $\eta$ = 0.2 in
the calculation according to ~\citet{Guetta01} and
\citet{Molinari07}. The densities $n= 1 ~\rm cm^{-3}$ in ISM case
is assumed.
Since $\Gamma_0$ is weakly dependent on the unknown density of the
cicumburst medium in the ISM case, it can be measured relatively
accurate. The results have been presented in Tab.2. Usually we
have $\Gamma_0\sim {\rm a~few \times 100}$ in the case of ISM with
a number density $n\sim 1~{\rm cm^{-3}}$, consistent with some
other independent probes
\citep{Sari99,Lithwick01,Zhang06,Molinari07,Jin07,Peer07,ZP08,ZP09a},
some of which are independent of the profile of the circumburst
medium.



\section{Conclusion}
We use the current {\it Swift}-XRT data to constrain the profile of
the circumburst medium and then measure the initial Lorentz factor
of the fireballs. As a reliable probe,  the X-ray light curves
should have the following characters for our purpose: (i) There is a
distinguished peak. (ii) Across the peak, a smooth transition to a
single power-law flux is followed, and the whole profile must be
consistent with the standard afterglow model. The early peaks
accompanying with steep decay, usually steeper than $t^{-3}$, are
abandoned. Among the $\sim 160$ {\it Swift} bursts we have checked,
only $4$ events meet such requests because in most events the early
X-ray emission are polluted by the emission powered by the prolonged
activity of the central engine.  In all these 4 bursts, the initial
increase of the X-ray flux is quicker than $t^2$, strongly suggests
a constant low-density medium. In general, we find {\bf $\Gamma_0
\sim {\rm a~few~\times 100}$}, consistent with the constraints
obtained in other analysis. Please note that in this work all the
events in the sample are bright GRBs. For the nearby sub-luminous
events, like GRB 980425, GRB 031203, GRB 060218 and GRB 060614, a
reliable estimate of their initial Lorentz factors are still
infeasible at present.

In our analysis we did not find a good case for the burst born in
stellar wind and behaving as a thin shell (see section 3.2). One
possible reason is that usually the wind medium is so dense that
the outflow has got decelerated significantly in a timescale
$t<T_{90}$. So the X-ray data is likely to be not suitable as a
probe of the initial Lorentz factor of GRB outflows for most of
events occurring in the wind medium.

{\it Note added in manuscript.}--After the acceptance of the paper,
the details of the X-ray afterglow data of GRB 090113 became
available (Krimm et al. 2009): The $0.3-10$ keV light curve shows an
initial period of roughly constant emission. For $t>530$ sec, the
light curve can be modelled with a power-law decay as $t^{-1.3}$.
The X-ray spectrum is $F_\nu \propto \nu^{-1.28\pm 0.20}$. One can
see that with  $p\sim 2.4$, both the temporal and the spectral
behaviors of this X-ray afterglow are well consistent with the
forward shock emission model in the case of a wind medium and
$\nu_{_{\rm X}}>\max \{\nu_c^{\rm f}, \nu_m^{\rm f}\}$. The initial
Lorentz factor can be estimated as
$\Gamma_0=[\frac{2E_{\gamma}(1+z)}{\pi A m_p c^3\eta t_p}]^{1/4}$,
where $A$ is related to the regular wind parameter $A_*$ by
$A=3\times 10^{35}A_*$. Assuming $z\sim 1.0$,  we have
$E_{\gamma}\sim7.9\times 10^{51} $erg and $\Gamma_0 \approx
121A_{34}^{-1/4}$.

\section{Acknowledgments}
We thank an anonymous referee for helpful comments and Nathaniel R.
Butler for providing us some data used in this work. This work made
use of data supplied by the UK Swift Science Data Centre at the
University of Leicester. This work is supported by the National
Science Foundation (grants 10673034 and 10621303) and National Basic
Research Program (973 programs 2007CB815404 and 2009CB824800) of
China. YZF is also supported by Danish National Research Foundation
and by Chinese Academy of Sciences.

\clearpage
\begin{table}
\begin{minipage}{170mm}
\caption{GRB sample}
\begin{tabular}{lccccccl}
\hline \hline
GRB& $z^a$ & $E_{\gamma}$($10^{52}$erg)$^b$ & $\rm T_{90}(s)^c$  &$\rm t_p(s)$ & medium type & $\Gamma_0^d$ & references\\
\hline
060801  &1.131 &0.55 &0.5 &110  &ISM &280  &1,2,3 \\
060926  &3.208 &1.0 &8.0 &430  &ISM  & 234& 4,2,5\\
080319C &1.95  &6.13 &34 & 360   &ISM &274  &6,7\\
080413B &1.1   & 1.53  &8.0 & 167 &ISM & 271 &8,9 \\
090113  &      & 0.79$^e$  &9.1 & 523 &wind & 121$A_{34}^{-1/4}$ &10 \\
\hline
\end{tabular}\\
References: 1 \citet{Cucchiara06}; 2 \citet{Butler07}; 3
\citet{Sato06}; 4 \citet{Elia06}; 5 \citet{Cummings06}; 6
\citet{Wiersema08}; 7 \citet{Stama}; 8 \citet{Vreeswijk08}; 9
\citet{Barthelmy08}; 10 \citet{Tueller}
\\
$^a$ redshift of GRB \\
$^b$ isotropic equivalent energy of prompt $\gamma$-ray emission
(1-$10^4$keV in the burst frame for bursts with defined redshift
or 1-$10^4$ keV in the observer frame for bursts with undefined
redshift
to consistent with the catalog in \citet{Butler07})\\
$^c$ time duration of GRB\\
$^d$ the calculated initial Lorentz factor of fireballs (assuming
$n=1\rm cm^{-3}$ for ISM case)\\
$^e$ assuming z=1 to derive $E_{\gamma}$ for GRB 090113 with
undefined redshift
\end{minipage}
\end{table}


\begin{thebibliography}{}

\bibitem[Akerlof et al. (1999)]{Akerlof99} Akerlof, C., Balsano, R., Barthelmy, S., et al. 1999, Nature, 398, 400
\bibitem[Bhattacharya (2001)]{Baha01}
Bhattacharya, D. 2001, Bulletin of the Astronomical Society of
India, 29, 107
\bibitem[Barthelmy et al. (2008)]{Barthelmy08} Barthelmy, S. D., Baumgartner, W., Cummings, J. R., et
al. 2008, GCN Circ. 7606
\bibitem[Blandford \& Mckee (1976)]{BM76}Blandford, R. D. \& McKee, C. F. 1976, Phys. Fluids, 19, 1130
\bibitem[Butler et al. (2007)]{Butler07}Butler, N. R., Kocevski, D.,
Bloom, J. S., \& Curtis, J. L. 2007, \apj, 671, 656
\bibitem[Cheng \& Wei (1996)]{wc96} Cheng, K. S., \& Wei, D. M.
1996, MNRAS, 283, L133
\bibitem[Chevalier \& Li (2000)]{CL00} Chevalier, R. A. \&
Li, Z. Y. 2000, \apj, 536, 195
\bibitem[Costa et al. (1997)]{Costa97} Costa, E., et al. 1997, Nature, 387, 783
\bibitem[Cucchiara et al. (2006)]{Cucchiara06} Cucchiara, A., Fox, D.
B., Berger, E., \& Price, P. A. 2006, GCN Circ. 5470
\bibitem[Cummings et al. (2006)]{Cummings06} Cummings, J. R., Barbier, L., Barthelmy, S.
D. et al. 2006, GCN Circ. 5621
\bibitem[Dai \& Cheng (2001)]{DC01} Dai, Z. G. \& Cheng, K. S. 2001, ApJ, 558,
L109
\bibitem[D'Elia et al. (2006)]{Elia06} D'Elia, V., Piranomonte,
S., Covino, S., Malesani, D., Fiore, F., Antonelli, A., Tagliaferri,
G., Stella, L., \& Chincarini, G. 2006, GCN Circ. 5637
\bibitem[Evans et al. (2007)]{Evan07} Evans, P. A., Beardmore, A. P., Page, K. L. et al. 2007, A\&A, 469, 379
\bibitem[Falcone et al. (2007)]{Fal08} Falcone, A. D., et al., 2007, ApJ, 671, 1921
\bibitem[Fan et al. (2002)]{Fan02} Fan, Y. Z., Dai, Z. G., Huang, Y. F., \& Lu, T.
2002, ChJAA, 2, 449
\bibitem[Fan \& Wei (2005)]{Fan05} Fan, Y. Z. \& Wei, D. M.
2005, MNRAS, 364, L42
\bibitem[Gallant (2002)]{Gallant02} Gallant, Y. A. 2002, Lecture
Notes in Physics., 589, 24
\bibitem[Gehrels et al. (2004)]{Gehrels04}Gehrels, N., et al. 2004, \apj,
611, 1005
\bibitem[Guetta, Spada \& Waxman (2001)]{Guetta01}Guetta, D., Spada, M. \& Waxman., E., 2001, \apj, 557, 399
\bibitem[Holland et al. (2006)]{Holland06} Holland, S. T.,
Barthelmy, S. D., Starling, R., Barbier, L. M., Perri, M., Capalbi,
M., Roming, P., Page, K., Nousek, J., \& Gehrels, N. 2006, GCN
Report 1.2
\bibitem[Holland et al. (2008)]{Holland08} Holland, S. T., Palmer,
D. M., Starling, R. L. C., \& Schady, P. 2008, GCN Report 128.1
\bibitem[Jin \& Fan (2007)]{Jin07} Jin, Z. P. \& Fan, Y. Z. 2007,
MNRAS, 378, 1043
\bibitem[Klebesadel et al. (1973)]{Kle73} Klebesadel, R. W., Strong, I. B. \& Olson, R.
A. 1973, \apj, 182, L85
\bibitem[Kobayashi (2000)]{Kob00}Kobayashi, S., 2000, ApJ, 545, 807
\bibitem[Krimm et al. (2009)]{Krimm}Krimm, H. A., Evans, P. A., Oates, S. R., et al.
2009, GCN Report 193.1
\bibitem[Lithwick \& Sari (2001)]{Lithwick01} Lithwick, Y. \& Sari, R. 2001, \apj, 555, 540
\bibitem[M\'esz\'aros \& Rees (1999)]{MR99} M\'esz\'aros P., and Rees M. J., 1999, MNRAS, {306}, L39
\bibitem[Molinari et al. (2007)]{Molinari07} Molinari, E., Vergani,
S. D., Malesani, D. et al. 2007, A\&A, 469, L13
\bibitem[Oates et al. (2009)]{Oates09} Oates, S. R., Page, M. J.,
Schady, P., et al. 2009, MNRAS accepted (astro-ph/0901.3597)
\bibitem[Pagani et al. (2008)]{Pagani08} Pagani, C., Racusin, Kennea, J.A. et al. 2008, GCN Circ
7460
\bibitem[Page et al. (2008)]{Page08} Page, K. L., Osborne, J. P., \&
Holland, S. T. 2008, GCN Circ 7376
\bibitem[Pe'er et al. (2007)]{Peer07} Pe'er, A., Ryde, F.,  Wijers,
R. A. M. J., M\'esz\'aros, P., \& Rees, M. J. 2007, ApJ, 664, L1
\bibitem[Piran, Shemi \& Narayan (1993)]{Piran93} Piran, T., Shemi, A., \& Narayan., R. 1993, MNRAS, 263, 681
\bibitem[Racusin et al. (2006a)]{Racusin06a}Racusin, J. L., Barbier,
L. M., Brown, P. J., Burrows, D. N.,  Gehrels, N., Guidorzi, C.,
Kennea, J. A., Mangano, V., Marshall, F. E., McLean, K. M.,
Palmer, D. M., Stamatikos, M., \& Troja, E. 2006a, GCN Circ. 5378
\bibitem[Racusin et al. (2006b)]{Racusin06b}Racusin, J. L., Grupe,
D., Morris, D., \& Stroh, M. 2006b, GCN Circ. 5382
\bibitem[Racusin et al.(2008)]{Racusin08}Racusin J. L., et al., 2008, Nature,
455, 183
\bibitem[Rhoads (1999)]{Rhoads99} Rhoads, J. E., 1999, ApJ, 525, 737
\bibitem[Sari et al. (1998)]{SPN98} Sari, R., Piran, T., \& Narayan, R. 1998,
\apj, 497, L17
\bibitem[Sari \& Piran (1999)]{Sari99} Sari, R., \& Piran, T. 1999,
\apj, 520, 641
\bibitem[Sato et al. (2006)]{Sato06} Sato, G., Barbier, L., Barthelmy, S. D. et
al. 2006, GCN Circ. 5381
\bibitem[Shemi \& Piran (1990)]{SP90} Shemi, A., \& Piran, T. 1990, \apj,
365, L55
\bibitem[Stamatikos et al. (2008a)]{Stama} Stamatikos, M., Barthelmy, S. D., Cummings,
J. et al. 2008a, GCN Circ. 7483
\bibitem[Stamatikos et al. (2008b)]{Stamaticko08} Stamatikos, M., Beardmore, A. P., Gehrels, N. et al.
2008b, GCN Circ. 7598
\bibitem[Troja \& Stamatikos (2008)]{Troja08}Troja, E. \&
Stamatikos, M. 2008, GCN Circ 7608
\bibitem[Tueller et al. (2009)]{Tueller} Tueller, J., Barthelmy, S. D., Baumgartner, W. H., et al. 2009, GCN
Circ. 8808
\bibitem[Vreeswijk et al. (2008)]{Vreeswijk08}Vreeswijk, P. M., Thoene, C. C., Malesani,
D. et al. 2008, GCN Circ. 7601
\bibitem[Waxman (1997)]{Waxman97} Waxman, E. 1997, \apj, 485, L5
\bibitem[Wei (2008)]{Wei07} Wei, D. M. 2007, MNRAS, 374, 525
\bibitem[Wiersema et al. (2008)]{Wiersema08}Wiersema, K., Tanvir,
N., Vreeswijk, P., Fynbo, J., Starling, R., Rol, E., \& Jakobsson,
P. 2008, GCN Circ 7517
\bibitem[Wijers, Rees \& M\'esz\'aros (1997)]{Wijers97} Wijers, R. A., Rees, M. J. \& M\'{e}sz\'{a}ros, P. 1997, MNRAS, 288, L51
\bibitem[Zhang et al. (2006)]{Zhang06} Zhang, B., Fan, Y. Z., \& Dyks, J. et al.
2006, ApJ, 642, 354
\bibitem[Zhang (2007)]{Zhang07} Zhang, B., 2007, Chin. J. Astron. Astrophys., 7,
1
\bibitem[Zou et al. (2005)]{Zou05} Zou, Y. C., Wu, X. F., \& Dai, Z.
G. 2005, MNRAS, 363, 93
\bibitem[Zou et al. (2009)]{ZP09a} Zou, Y. C., Fan, Y. Z., \& Piran. T., 2009,
MNRAS submitted (astro-ph/0811.2997)
\bibitem[Zou \& Piran (2009)]{ZP08} Zou, Y. C., \& Piran. T., 2009, in
preparation




\end{thebibliography}
\end{document}